\documentclass[twocolumn,prb,citeautoscript,showpacs]{revtex4-1}

\usepackage{amsmath}
\usepackage{amssymb}
\usepackage{graphicx}
\usepackage{hyperref}

\begin{document}

\title{Electron doping in $\text{Sr}_3\text{Ir}_2\text{O}_7$: collapse of band gap and magnetic order}

\author{Michael W. Swift}
\affiliation{Department of Physics, University of California, Santa Barbara, California 93106-9530, USA}

\author{Zach Porter}
\affiliation{Department of Physics, University of California, Santa Barbara, California 93106-9530, USA}

\author{Stephen D. Wilson}
\affiliation{Materials Department, University of California, Santa Barbara, California 93106-5050, USA}

\author{Chris G. Van de Walle}
\affiliation{Materials Department, University of California, Santa Barbara, California 93106-5050, USA}

\begin{abstract}
The electron-doping-driven collapse of the charge gap and staggered magnetization of the spin-orbit-assisted Mott insulator Sr$_{3}$Ir$_{2}$O$_{7}$ is explored via first-principles computational methods.  In the antiferromagnetic phase, the gap and magnetization are observed to decrease slowly with increasing doping, with an abrupt collapse of both the gap and the magnetization at an electron concentration corresponding to 4.8\% substitution of Sr with La, in excellent agreement with experiment.  Additionally, we describe the structural effects of electron doping in Sr$_{3}$Ir$_{2}$O$_{7}$ via a competition between the steric effect from smaller La atoms substituted within the lattice and the dominant doping-driven deformation-potential effect. Curiously, our first-principles calculations fail to capture the low-temperature structural distortion reported in the low-gap phase of Sr$_{3}$Ir$_{2}$O$_{7}$, supporting the notion that this distortion arises as a secondary manifestation of an unconventional electronic order parameter in this material.
\end{abstract}

\maketitle

%\section{Introduction}
%\label{intro}

In the Ruddlesden-Popper iridates (Sr$_{n+1}$Ir$_{n}$O$_{3n+1}$), the correlation energy $U$ is comparable in magnitude to the spin-orbit coupling energy $\lambda$.  This gives rise to an unusual variation on typical Mott insulating behavior in which a combination of $U$ and $\lambda$ opens a charge gap.~\cite{Kim08,Moon08,Arita12}  Iridium in these compounds is in the $4+$ oxidation state with the $t_{2g}$ states occupied with 5 electrons; however in both the $n=2$ case (Sr$_3$Ir$_2$O$_7$, or Sr-327) and the $n=1$ case (Sr$_2$IrO$_4$, or Sr-214), spin-orbit coupling entangles these states and lifts their orbital degeneracy, resulting in a fully occupied $J_\text{eff}=3/2$ quadruplet and a half-occupied $J_\text{eff}=1/2$ doublet.  This splitting, combined with on-site Coulomb repulsion, opens a Mott-Hubbard gap in the half-occupied doublet, forming an insulating antiferromagnetic (AF) $J_\text{eff}=1/2$ ground state. The spin-orbit-assisted stabilization of a Mott gap amongst $J_\text{eff}=1/2$ electrons renders these materials excellent platforms for the observation of many exotic phenomena including proposals of high-temperature superconductivity \cite{Wang11}.

The metallic state realized just beyond the antiferromagnetic insulating $J_\text{eff}=1/2$ Mott state is often an anomalous one.  The propensity for these weakly correlated spin-orbit assisted Mott states to host unconventional metallic phases is rapidly being established.  Examples include the formation of pseudogapped metals \cite{Kim14,delaTorre15,Cao16} with proposals of incipient superconductivity \cite{Yan15,Kim16}, the formation of spin density wave metals~\cite{Dhital14} and paramagnetic states with unconventional spin dynamics~\cite{Hogan16, Kim12, Vale15}, the emergence of competing electronic order parameters~\cite{Hogan15, Hsieh12, Wang13, Chu17}, two-dimensional metallic phases \cite{PhysRevLett.116.216402}, and reports of negative electronic compressibility~\cite{He15}.  Understanding the underlying mechanisms driving many of these exotic phase phenomena remains an ongoing challenge.

Of particular interest is the metal-insulator transition (MIT) obtained upon electron doping in Sr-327.  Through substitution of lanthanum on the strontium site [forming (Sr$_{1-x}$La$_{x})_3$Ir$_2$O$_7$], experiments have observed a MIT near a lanthanum concentration $x=0.04$~\cite{Hogan15,Chu17}.  This transition is noteworthy because of a number of open questions regarding its nature and driving mechanisms.  In this paper, we seek to advance the understanding of these questions through the use of first-principles calculations modeling the response of Sr-327 to electron doping.  While previous theoretical works have discussed the band structure~\cite{Kim08,Moon08,Arita12,Liu15,Kim17_2}, crystal structure~\cite{Hogan16}, and magnetic order~\cite{Boseggia12,Kim17} of the Ruddlesden-Popper iridates, first-principles calculations directly addressing electron doping are notably absent.  Our results provide a step forward in understanding the MIT that occurs in electron-doped Sr-327.

The first question we will address is the nature of the experimentally observed coupling between the charge gap and the onset of antiferromagnetic order below the transition~\cite{Chu17,Hogan15}.  We calculate an abrupt collapse of the gap and the staggered magnetization at an electron concentration corresponding to 4.8\% substitution of strontium with lanthanum, consistent with experimental results.  We will argue that the antiferromagnetic order is essential to the stabilization of the gap, and works in combination with the Mott-Hubbard interactions to form the insulating state.  

Another open question concerns the shifts in lattice parameters upon lanthanum doping.~\cite{Hogan15,Hogan17}  We present new experimental data on these shifts.  In our calculations, we separate the effects of ionic substitution from electron doping.  We find that ionic substitution leads to a steric tendency to contract the lattice, which competes with the electronic tendency to expand the lattice.

It is also worth noting that the paramagnetism in La-doped Sr-327 is not well understood.  The paramagnetic state is not that of a conventional bulk paramagnet but rather one of defect driven local moments that arise via La substitution, with $J=1/2$ impurity moments being induced for each La cation substituted~\cite{Hogan16_2}.  While interesting, this question is not well suited to our theoretical methodology.  Since local moments are only present in a small volume fraction of the sample, they are unlikely to materially impact the bulk metallic state in any case and thus will not be discussed further here.

The final set of questions we will address concerns the charge-density-wave (CDW)-like instability and coincident structural distortion that arise in the metallic state of La-doped Sr-327~\cite{Chu17}.  While some have suggested that phase competition from this electronic order parameter is the driver of the MIT, we will argue that this is not the case.  Instead, these phenomena necessarily arise from many-body correlation physics active in the metallic regime separate from the interactions which stabilize the gap.

%\section{Computational Methods}
%\label{methods}

Our calculations are based on density functional theory (DFT), performed using the {\it Vienna Ab-initio Simulation Package} (VASP)~\cite{VASP} with the projector-augmented-wave method~\cite{PAW}.  Spin-orbit coupling is included using the non-collinear spinor method as implemented in VASP.  The spin quantization axis is chosen along an in-plane lattice vector; tests show that results are insensitive to this choice.  Correlation effects are taken into account using the DFT+$U$ approach~\cite{Dudarev98} with the Perdew-Burke-Ernzerhof (PBE) functional~\cite{PBE}.  The value of the on-site Hubbard $U$ parameter is taken to be $U=1.6$ eV based on constrained random-phase-approximation calculations from Ref.~\citenum{Kim17}. 

Experimental results for lattice parameters of doped samples at 300 K are extracted from Le Bail refinements of laboratory-source x-ray diffraction (XRD) profiles to the structural model in Ref.~\citenum{Hogan16}. Lanthanum and calcium substitution levels for the samples were measured using wavelength-dispersive x-ray fluorescence spectroscopy (WDXRF), in addition to energy-dispersive x-ray spectroscopy (EDS) to confirm microscopic uniformity.

The starting structure in the calculations is taken from the 100-K refinement in Ref.~\citenum{Hogan16}, and the lattice parameters and ionic positions are fully relaxed in all calculations.  The Brillouin zone is sampled using a $4\times4\times1$ $\Gamma$-centered grid; only a single {\bf k}-point point is needed in the $k_z$ direction because of the size of the $c$ lattice parameter and because of the quasi-2D nature of the material.  To capture the details of the conduction band with sufficient accuracy, the zone is sampled more finely near the conduction-band minimum (CBM).  The Voronoi cell of the $k$ point closest to the CBM is sampled with a fine mesh, with density equivalent to a $24\times 24 \times 1$ grid.  Supercells used for studying lanthanum and calcium substitution contain $2\times 2\times 1$ unit cells and use a $2\times2\times1$ $\Gamma$-centered $k$ point grid.  All calculations use a plane-wave energy cutoff of 500 eV.

%\section{Results and Discussion}

%\subsection{Electron Doping Sr-327}
%\label{doping}

\begin{figure}
\includegraphics[width=\columnwidth]{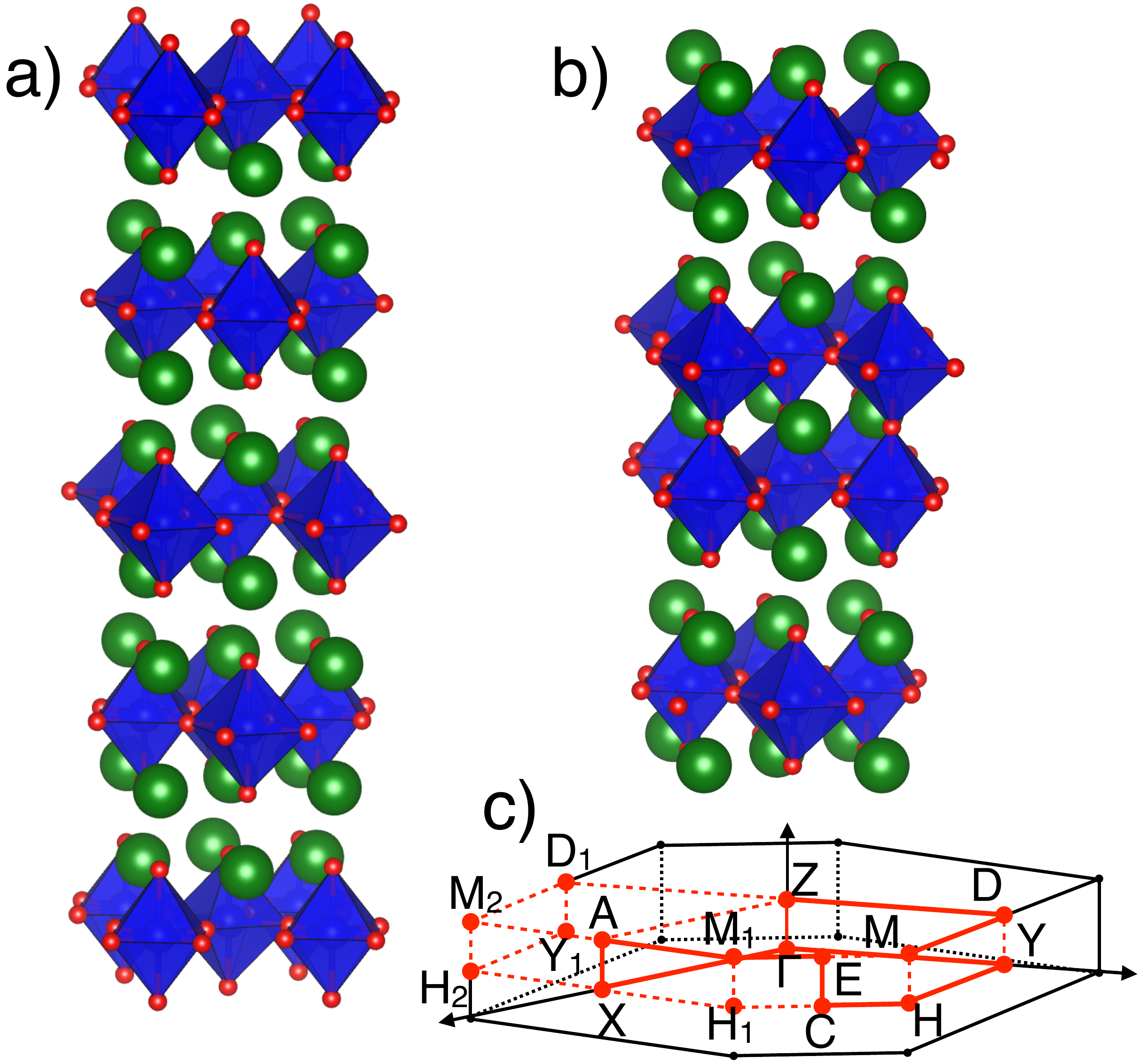}
\caption{Structural visualizations of a) Sr$_2$IrO$_4$ and b) Sr$_3$Ir$_2$O$_7$.  Sr-214 takes a tetragonal structure (space group $I4_1/acd$)~\cite{PhysRevB.49.9198}, and slight deviations from tetragonal make Sr-327 monoclinic (space group $C2/c$)~\cite{Hogan16}.  The first Brillouin zone for Sr-327 is shown in c), with high-symmetry points labeled~\cite{Setyawan10}. }
\label{fig:struct}
\end{figure}

\begin{figure}
\includegraphics[width=\columnwidth]{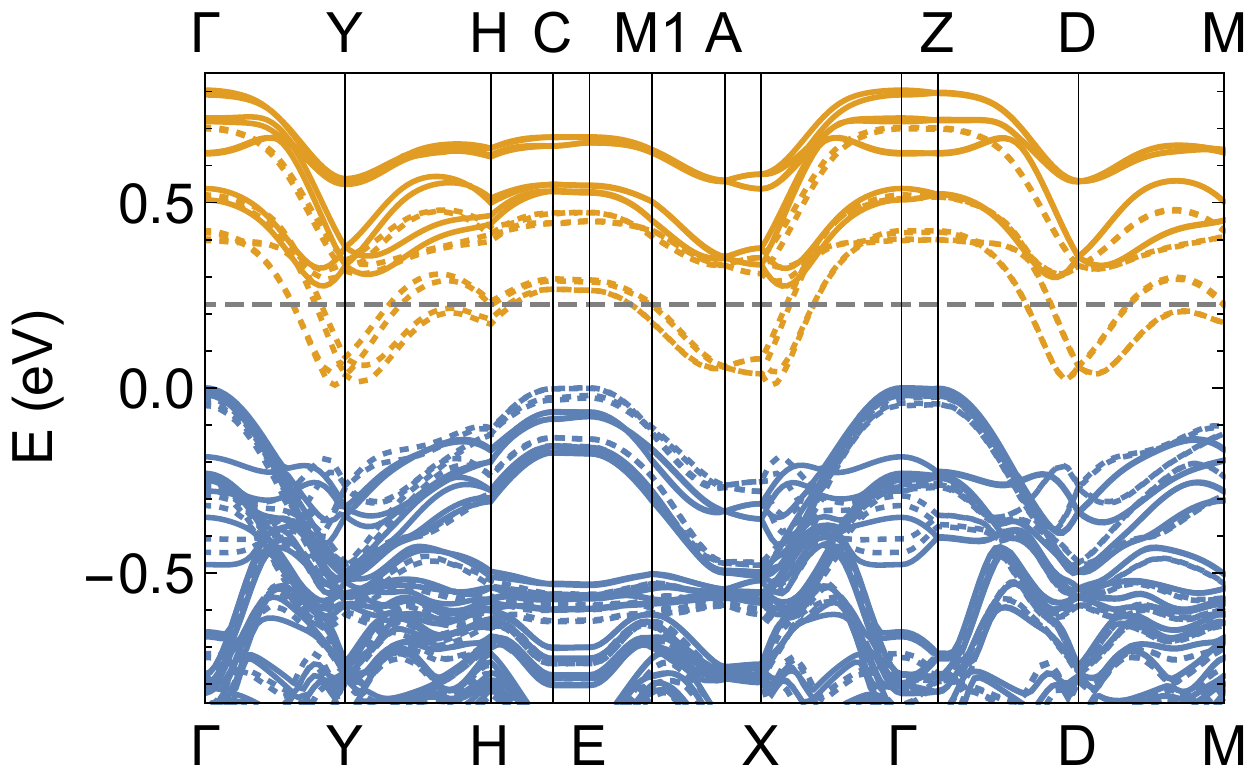}
\caption{DFT+$U$ band structure of Sr-327.  Solid lines show the undoped (insulating) case, and dashed lines show the band structure after the collapse of the gap (at a doping level $x=0.08$).  The energy of the bands is referenced to the VBM in both cases.  Occupied states (Ir $5d$ states with a mix of $J_\text{eff}=3/2$ and $J_\text{eff}=1/2$) are shown in blue, and unoccupied states (Ir $5d$ states with $J_\text{eff}=1/2$) are shown in orange.  The gap in the undoped case is 0.27 eV, indirect between $\Gamma$ and a point 81\% of the way from $\Gamma$ to X.  The smallest direct gap is 0.44 eV, 78\% of the way from $\Gamma$ to X.  In the doped case, the indirect gap has collapsed to only 7 meV.  The CBM is at the same point of the Brillouin zone as in the undoped case, and the VBM is at the C point.  Also shown is the Fermi level in the doped case, represented by a gray dashed line 218 meV above the CBM. }
\label{fig:BS}
\end{figure}

The Ruddlesden-Popper iridates ($n=1$ and $n=2$) are formed of $n$ layers of corner-sharing IrO$_6$ octahedra with Sr in the voids, interspersed with SrO rock-salt layers.~\cite{Moon08}  Visualizations of the structures are given in Fig.~\ref{fig:struct}, and the band structure of Sr-327 is shown in Fig.~\ref{fig:BS}.  The lattice structure of Sr-327 was recently identified to have a subtly monoclinic C2/c symmetry \cite{Hogan16} and the system is known to order antiferromagnetically below 280 K \cite{Matsuhata04,Chu17} with an ordered moment of $m\approx\frac{1}{3}$ $\mu_B$ \cite{Dhital13,Liu15,Hogan15}.

Initially, the magnitude of the charge gap of Sr-327 was difficult to experimentally access, with reported values ranging from 85 meV from optical spectroscopy~\cite{Moon08,Park14} to 0.3 eV from angle-resolved photoemission spectroscopy (ARPES) data on doped samples~\cite{delaTorre14,He15}. This suggests the optical spectroscopy (as well as DC transport) measurements may probe a sub-band-gap or phonon-assisted transition \cite{PhysRevB.89.155115} rather than the gap itself.  In a number of other Mott insulators, defects or small polarons have been identified as sources of sub-band-gap transitions observed in optical spectroscopy~\cite{Himmetoglu14,Bjaalie16}. Our calculated value of the band gap, 0.27 eV, agrees well with reported ARPES data and is consistent with the notion of polaronic effects in Sr-327 \cite{PhysRevB.87.241106}.

In experiments, doping is typically accomplished by substituting some fraction of the strontium with lanthanum, which acts as an electron donor.~\cite{Li13,delaTorre14,Chen15,He15,Hogan15,Hogan16_2,Chu17}  The effects of cation substitution and electron doping are thus inextricably linked.  The bare effect of cation substitution may be approximated by substituting calcium, since it is isoelectronic with strontium but similar in size to lanthanum,~\cite{Hogan15} but this approximation is imperfect.  First-principles calculations can explicitly separate the cationic and electronic effects.

To examine pure electron doping, we introduce fractional electrons to the unit cell, with overall charge neutrality ensured by a uniform compensating background.  This allows us to capture the physics of electron doping in isolation from the effects of lanthanum impurities that introduce electrons in an experiment.  As the number of electrons in the cell increases, there is a slight decreasing trend in the gap and the magnetization.  We are able to stabilize two different states.  The ground state has a 0.27 eV gap and shows significant antiferromagnetic ordering with staggered magnetization 0.29 $\mu_B$ aligned out-of-plane.  The higher-energy state has a smaller gap (0.11 eV) and shows a much smaller canted aniferromagnetic ordering with staggered magnetization 0.06 $\mu_B$.  This is consistent with the experimentally observed enhancement of the gap in the antiferromagnetic state~\cite{Hogan15}.

\begin{figure}
\includegraphics[width=\columnwidth]{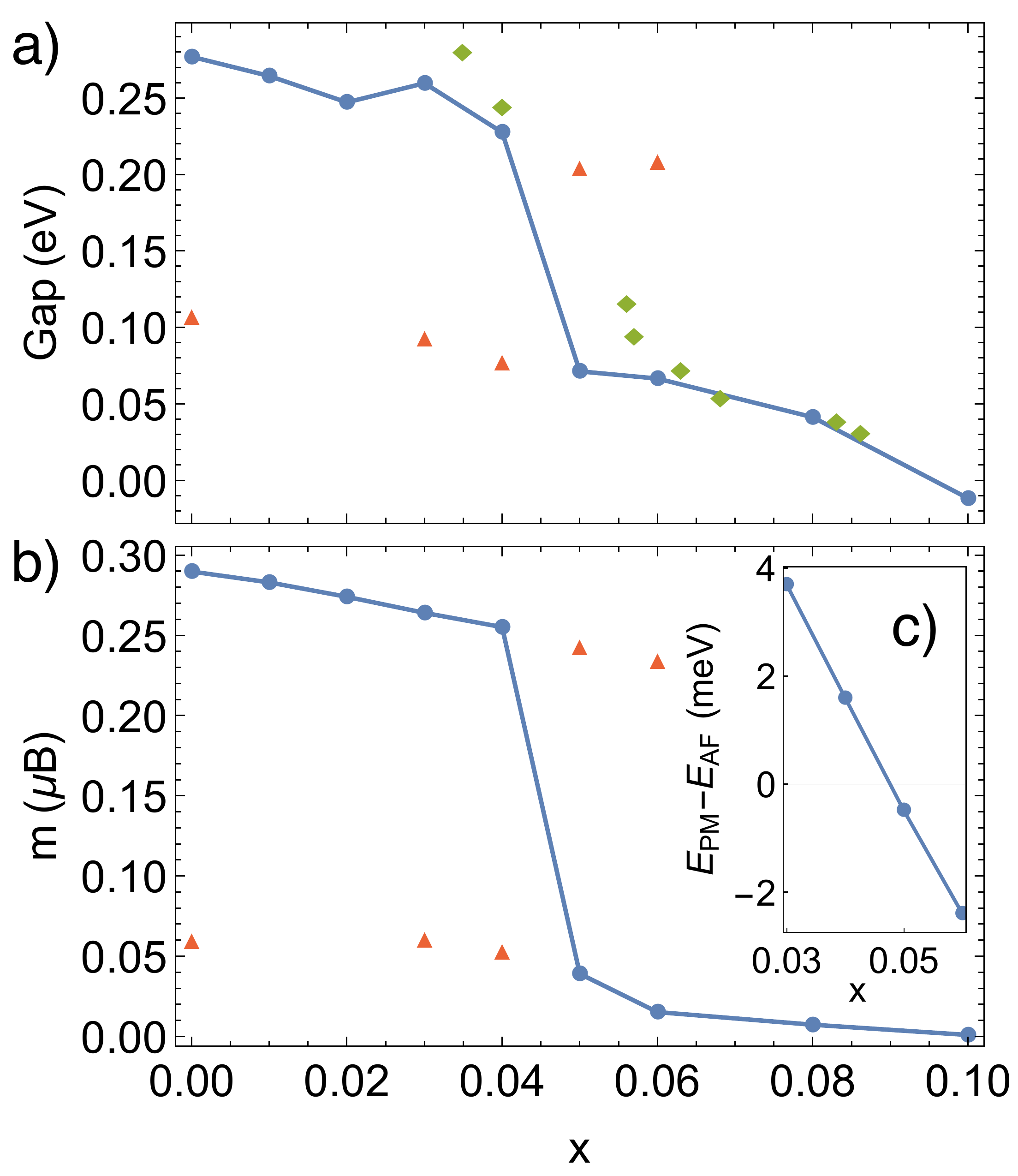}
\caption{a) Indirect band gap and b) out-of-plane magnetization as a function of the number of doping electrons per strontium atom ($x$).  Lowest-energy structures for each doping level are shown in blue circles, connected by lines as a guide to the eye.  Calculated metastable structures are shown in red triangles.  In panel a), experimental gap values from Ref.~\citenum{He15} are shown as green diamonds for comparison.  Inset c): energy difference between the small-moment state and antiferromagnetic (AF) state.  The crossover at $x=0.048$ is the critical doping level for the phase transition.   }
\label{fig:gap}
\end{figure}

At each doping level the structure with the lower total energy was identified as the ground state.  As seen in Fig.~\ref{fig:gap}, at low doping the antiferromagnetic state is more stable, but the smaller-gap state becomes increasingly favorable as doping increases.  At $x=0.048$, or 0.144 electrons per formula unit, the small-moment state becomes the favored ground state.  This causes the gap and magnetization to collapse simultaneously.  The calculated magnitude of the sublattice magnetization and its evolution with doping are in good agreement with experiments, and the critical value for the transition is close to the observed $x=0.04$ for the metal-insulator transition.~\cite{Hogan15}   The collapse of the gap is in excellent agreement with direct measurements of the band gap in doped samples using ARPES.~\cite{He15}  These results suggest that the calculated phase transition from the high-gap to the low-gap state is the cause of the insulator-metal phase transition observed in experiments.

It is significant that the collapse of the charge gap is associated with the collapse in magnetization, even in the metastable states.  This is strong evidence that these two phenomena are fundamentally related, and the magnetic order may play a role in stabilizing the gap.

The shape of the conduction and valence bands remains fairly constant with doping; the collapse in the band gap is due primarily to an overall shift of the bands, with changes in band shape playing only a minor role.  The CBM remains at the valley 81\% of the way from $\Gamma$ to X for all values of doping ($x=0.0$ to $0.10$).  The valence band has two local maxima, one at $\Gamma$ and the other at C.  At $x=0$ the valence-band maximum (VBM) is at $\Gamma$; C is lower by 52 meV.  As doping increases, the extremum at C rises, until it passes $\Gamma$ to become the VBM beyond $x=0.06$.  At $x=0.10$, the highest level of doping we tested, the $\Gamma$ maximum is lower than the C maximum by 38 meV.  See Fig.~\ref{fig:BS}.

In the simplest spin-orbit Mott picture, the character of the upper and lower Mott-Hubbard band is iridium $d$.  This is indeed the predominant character of the calculated states, but there is strong hybridization with oxygen $p$ states, as shown in Table~\ref{table:hybridization}.  This hybridization varies across the Brillouin zone, but is remarkably consistent across the high-gap AFM state and the low-gap small-moment state, and changes little with electron doping.
This strong hybridization is consistent with earlier work~\cite{Moon08}, explaining observed magnetic moments on the oxygen sites~\cite{Miyazaki15} and various signatures in optical spectroscopy~\cite{Park14}.  Hybridization with oxygen states also plays a role in the dimensionality-induced phase transition across the Ruddlesden-Popper series.~\cite{Kim17_2}

\begin{table}
\begin{tabular}{lll}
\multicolumn{3}{l}{High-gap AFM state, undoped}\\
\hline
State & Ir $d$ & O $p$ \\
VBM & 51\% & 36\% \\
Valence band C point & 54\% & 32\%\\
CBM & 61\% & 28\% \\ 
\hline \\
\multicolumn{3}{l}{Low-gap small-moment state, undoped}\\
\hline
State & Ir $d$ & O $p$ \\
VBM & 51\% & 36\% \\
Valence band C point & 56\% & 30\% \\
CBM & 61\% & 28\% \\ 
\hline \\
\multicolumn{3}{l}{Low-gap small-moment state, $x=0.1$}\\
\hline
State & Ir $d$ & O $p$ \\
VBM & 52\% & 35\% \\
Valence band C point & 59\% & 27\% \\
CBM & 60\% & 29\% \\\hline
\end{tabular}
\caption{Fraction of the charge density of various states projected onto iridium $d$ and oxygen $p$ states for the high-gap AFM state at zero doping, the low-gap small-moment state at zero doping, and the low-gap small-moment state at $x=0.1$. }
\label{table:hybridization}
\end{table}

Recent experiments have observed a CDW instability in the metallic phase of Sr-327, accompanied by a lattice distortion that creates two inequivalent iridium sites with significantly modified octahedral tilting.~\cite{Chu17}  Though we allow for this symmetry breaking in our calculations, it does not occur, and when the distortion is introduced manually the calculations show it to be unstable.  It is significant that, while DFT is able to capture the collapse of the gap, the CDW and lattice distortions in the metallic state are not captured.  This suggests that these phenomena emerge from many-body effects in the correlated metal distinct from the mechanism behind the transition itself.  
More work is needed to explore the metallic state of this material and determine the source of the CDW-like instability.

%\subsection{Impurity Substitution and Lattice Parameters}
%\label{impurities}

\begin{figure}
\includegraphics[width=\columnwidth]{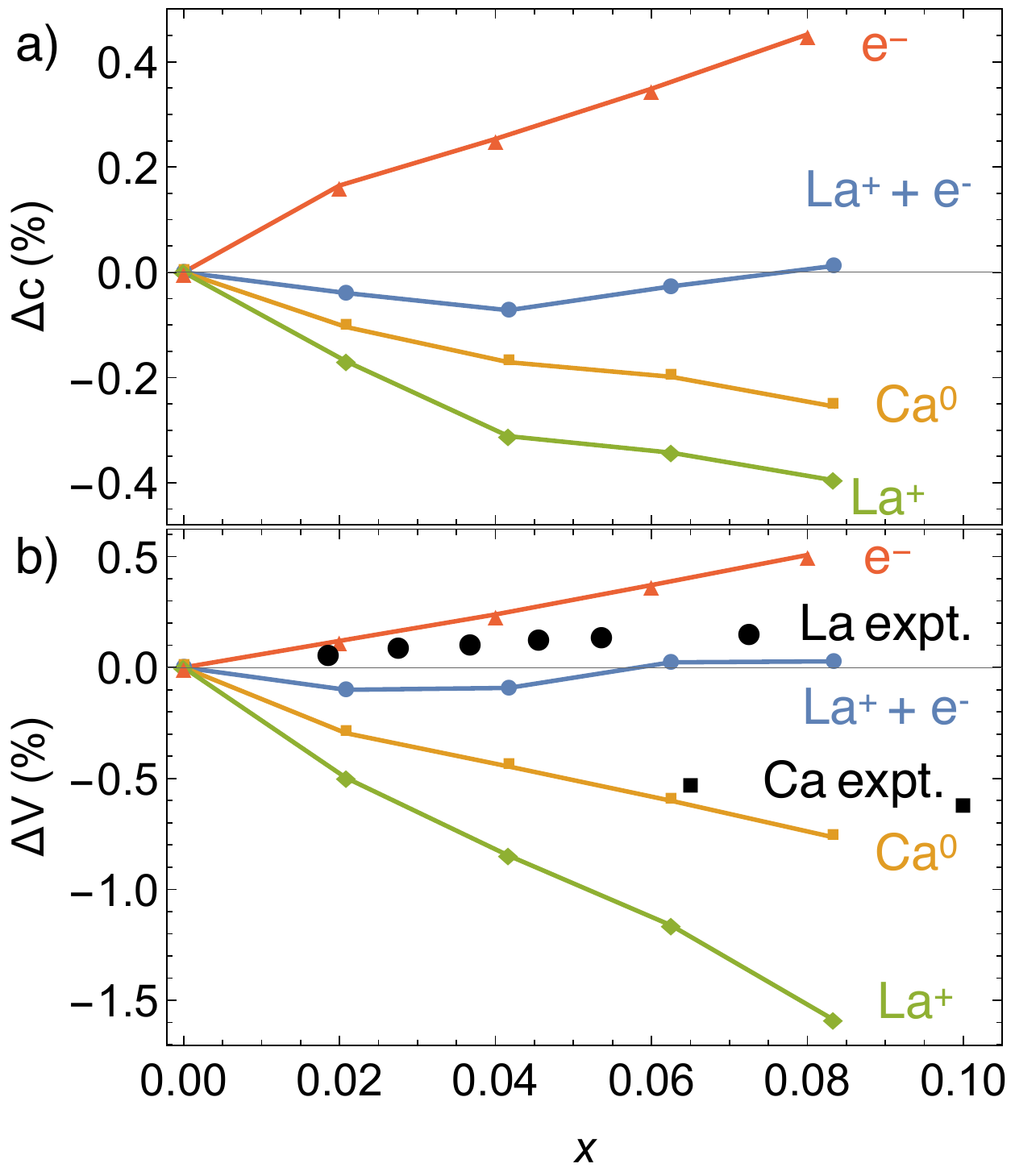}
\caption{Percentage change of a) out-of-plane lattice parameter $c$ and b) cell volume as a function of doping.  Note the tendency of the lattice to contract on calcium doping (or incorporation of La$^+$), driven by steric effects, and the tendency toward lattice expansion on electron doping, driven by the deformation-potential effect.   Lines connecting data points are a guide to the eye.  Electron doping (without ionic substitution) is shown in red triangles, La$^+$ (lanthanum substitution without the donated electron) is shown in green diamonds, and La$^+ + e^-$ (lanthanum substitution including the donated electron) is shown in blue circles.  Experimental data for lanthanum doping~\cite{Hogan15,Hogan17} are shown as solid black circles.  For comparison, calculated results for calcium substitution are shown as orange squares, and experimental results as black squares.  }
\label{fig:lattice_params}
\end{figure}

We now consider the other aspect of doping: ionic substitution.
Our results show that the primary effect of cation substitution is steric.  Since the ionic radii of lanthanum and calcium are smaller than that of strontium, they are expected to shrink the lattice.  We replace one strontium ion with a lanthanum or calcium ion within a supercell that is large enough to minimize the interaction between the defect and its periodic images.  The steric effect of lanthanum substitution is isolated from the electronic effect by removing the electron that the donor would donate to the conduction band (i.e., placing the system in a positive charge state).  As in the electron doping study, overall charge neutrality is ensured by a uniform compensating background.  Calculations which include electron doping show the same MIT found in bulk cells doped purely with electrons, without any atomic substitution.  Figure~\ref{fig:lattice_params} shows the lattice parameter decreasing with increasing incorporation of calcium and positively charged lanthanum, confirming the expectation based on ionic size.

Figure~\ref{fig:lattice_params} also shows the pure electronic effect of doping, which is a tendency toward lattice expansion.  
%Electron doping is calculated as described in section~\ref{doping}.
The effect of electron doping on the lattice is well known in semiconductors and referred to as a ``deformation potential'' effect.
Deformation potentials describe the shift of the bands as a function of strain, and when the conduction band is occupied with electrons, the energy of the system can be lowered by a deformation that lowers the energy of the conduction band minimum.  This creates a driving force for the expansion or contraction of the lattice, depending on the sign of the deformation potential.~\cite{Vandewalle03}  This theory has been successfully applied to other complex oxides,~\cite{Janotti12} and we expect that it explains the effect of electron doping on the lattice in Sr-327.
When both the electronic and ionic effects are applied [through a supercell containing both the lanthanum donor(s) and the associated free electron(s)], the electronic effect largely cancels out the steric effect.  

Figure~\ref{fig:lattice_params}(b) includes a comparison with experiment.  Lanthanum doping results are taken from Refs.~\onlinecite{Hogan15,Hogan17}, calcium doping results were obtained using the methodology outlined above.
Our calculated results are consistent with measurements of the fractional volume change for both lanthanum- and calcium-substituted samples [Fig.~\ref{fig:lattice_params}(b)].
For lanthanum, a very small effect on volume is observed;
%Note that with increasing lanthanum doping, experiments see a decrease in out-of-plane lattice parameter followed by an increase, consistent with our calculations.  
for calcium, our experimental results show a sizeable decrease in volume with increasing calcium doping, again consistent with our calculations.

%\section{Conclusion}
%\label{conclusions}

In summary, our calculations show a simultaneous collapse of the charge gap and the staggered magnetization in (Sr$_{1-x}$La$_{x})_3$Ir$_2$O$_7$ at $x=0.048$, in good agreement with experimental results.  This transition from a high-gap AFM state to a low-gap small-moment state comes about purely through the effect of electronic doping.  Ionic effects of lanthanum doping are primarily steric.  The tendency toward lattice contraction driven by ionic size is counteracted by the tendency of the electrons to expand the lattice through the deformation potential of the conduction bands. Our calculations do not show the CDW instability nor lattice distortions observed in the metallic state of Sr-327.  This suggests that the CDW and accompanying lattice distortion in the metallic state arise from many-body effects not present in our calculations, whereas the collapse of the gap and magnetization both arise from physics that is well-described by DFT+$U$ with spin orbit coupling.

%\section{Acknowledgments}

We thank Tom Hogan for fruitful discussions.  This work was primarily supported by the MRSEC Program of the National Science Foundation under Award No.\ DMR-1121053.  Additional support was provided by NSF Award No. DMR-1505549 (Z.P.). 
Computational resources were provided by the Extreme Science and Engineering Discovery Environment (XSEDE), which is supported by NSF grant number ACI-1548562, and the Center for Scientific Computing from the CNSI, MRL: an NSF MRSEC (DMR-1121053) and NSF CNS-0960316.

\bibliography{Sr-327}

\end{document}